\documentclass[preprint]{aastex}

\begin{document}

\title{{\it Chandra} ACIS and {\it Newton} EPIC Observations of the
X-ray Luminous SN1978K in NGC 1313}

\author{Eric M. Schlegel, Albert Kong, Philip Kaaret, Rosanne DiStefano,
Steve Murray}

\affil{High Energy Astrophysics Division, Smithsonian Astrophysical
Observatory, Cambridge, MA 02138}


\begin{abstract}

We describe an observation of the X-ray luminous SN1978K in NGC 1313
using the ACIS detector on {\it Chandra} and an archival {\it Newton}
EPIC observation.  The models that provided good fits to the {\it
ASCA} SIS and GIS and the {\it ROSAT} PSPC spectra no longer do so for
the ACIS and EPIC spectra.  The best-fit model to the ACIS and EPIC
spectra are dual hot plasma models (vmekal); one component is soft (T
= 0.61$^{+0.04}_{-0.05}$ keV, 90\% errors) and the other is harder (T
= 3.16$^{+0.44}_{-0.40}$ keV).  For the varying abundances permitted
within the model, only the Si abundance of the soft component differs
from solar, with a value n$^{\rm Si}_{\rm SN78K}$~/~n$^{\rm Si}_{\rm
solar}$ = 3.20$^{+1.80}_{-1.90}$ (90\% errors).  From a ratio of the
low- and high-T model fits to the {\it Chandra} and {\it XMM-Newton}
spectra, we infer an exponent ${\rm n}$ of the ejecta density
distribution, ${\rho}_{\rm ejecta}$ $\propto$ r$^{\rm -n}$, of
$\sim$5.2 adopting a circumstellar matter distribution exponent of s =
2 (${\rho}_{\rm cs}$ $\propto$ r$^{\rm -s}$).  The 0.5-2 keV light
curve shows essentially no decline; the 2-10 keV light curve,
constructed only of the {\it ASCA}, {\it XMM-Newton}, and {\it
Chandra} observations, shows a drop of 1.5 from the {\it ASCA} epoch.
The hard band decline, together with the apparently enhanced Si
emission, signal the start of the X-ray decline of SN1978K.

\end{abstract}

\keywords{supernovae: individual (SN1978K)}

\section{Introduction}

X-ray emission may emerge at two stages in the evolution of a
supernova: near the time of the shock breakout and via circumstellar
interaction at a later date.  Three different processes generate
X-rays: from the prompt thermal burst; via Compton-scattered
${\gamma}$-rays from the synthesized matter's radioactive decay (from
which X-rays are emitted while the debris is optically thick to
${\gamma}$-rays); and from the interaction of the shock with
circumstellar material in the vicinity of the progenitor.  Only
SN1987A has been seen to emit X-rays definitely attributed to
Compton-scattered ${\gamma}$-rays (Dotani et al. 1987, Nature, 330,
230).

Two mechanisms exist for the circumstellar production of X-rays:
X-rays from the outgoing shock or from the reverse shock
\citep{Fran96}.  Below 10 keV, the reverse shock should be the
dominant emitter because the density behind the shock is higher, by a
factor of $\sim$4, than the swept-up circumstellar density behind the
forward shock.  If a cool shell forms because the reverse shock is
radiative, that shell absorbs X-rays below 10 keV and the observed
1-10 keV emission must be from the forward shock as discussed for
SN1993J \citep{Fran96}.

Other than SN1987A, few supernova have been followed in the X-ray
band.  The total current pool consists of $<$20 objects; of those,
only four supernovae have data sets consisting of more than four
observations (SN1978K, SN1993J, SN1998S, SN1999em; \citealt{S95,IL02}).
SN1978K is one of a small number of extremely luminous, X-ray emitting
supernovae \citep{S95,IL02}.  As X-ray observations are difficult to
obtain, each bright supernova contributes valuable knowledge to our
understanding of the X-ray emission from supernovae.

The X-ray emission of SN1978K was first uncovered in an observation
obtained with the {\it ROSAT} PSPC in 1992 \citep{Ryder93} (hereafter,
R93).  It had been detected optically during a spectrophotometric
survey of H II regions two years earlier and reported as a nova
\citep{DopRyd90} at that time.  It had also been detected as a
powerful radio source in 1982 but never followed up (R93).  As part of
the investigation spurred by the X-ray detection, examination of
archival optical plates uncovered a light curve that while crude was
sufficient to assign an explosion date near 1 June 1978 (R93).

To further study SN19798K, observations were obtained during the mid-
and late-1990s with {\it ASCA}, {\it Hubble Space Telescope}, {\it
ROSAT} (High Resolution Imager), {\it Australia Telescope Compact
Array} (ATCA), and the {\it Anglo-Australian Observatory} (AAO)
covering the X-ray, ultraviolet, optical, and radio bands
\citep{Schl99} (hereafter, S99).  An early {\it ASCA} spectrum and
{\it ROSAT} light curve were reported in \cite{Petre94} and
\cite{Schl96}, respectively.  The {\it ATCA} radio light curve showed
behavior typical of a circumstellar interaction, with a rapid turn-on
followed by a slow decline at all radio wavelengths (e.g.,
\citealt{Weiler96}; S99).  Optically, the {\it AAO} spectra of SN1978K
showed strong emission lines of the Balmer series as well as He I, [O
III], [Ne III], and [Fe II] (R93).  In the ultraviolet, the Mg II
doublet and [Ne IV] lines were detected using the {\it HST} Faint
Object Spectrograph (S99).  The X-ray light curve was consistent with
no decline throughout the decade (S99).  The {\it ASCA} spectra were
best-fit with absorbed, single-component continuum models; no evidence
for line emission existed in the data (\citealt{Petre94}, S99).

SN1978K is important because it is among the first known X-ray
emitting supernovae with luminosities above 10$^{39}$ erg s$^{-1}$.
It is also a nearby object and may be studied in greater detail than
more distant supernovae.

In this paper, we present the {\it Chandra} ACIS spectrum of SN1978K
and contrast it with the recent {\it XMM-Newton} and earlier {\it
ASCA} spectra.  Throughout we adopt a distance to NGC 1313 of
4.13$\pm$0.11 Mpc as determined by \cite{Mendez02}.

\section{Data}

{\it Chandra} observed NGC 1313 and its X-ray emitting sources on 2002
October 13 for 20592 sec with the ACIS detector \citep{Gar03}.  The
aimpoint of the detector fell on the back-illuminated S3 CCD.  We
extracted a light curve of the background, masking out all bright
sources detectable by eye, to search for any signatures of flaring
events known to affect back-illuminated CCDs \citep{Chan02}.  No
strong flares were detected but a few high points were eliminated,
reducing the good exposure time to 19902 sec.

We extracted the events at the location of SN1978K, about 5$'$.88
off-axis, in an aperture of radius 12$''$ which encloses $>$98\% of
the point spread function \citep{Chan02}.  The background was obtained
from an annulus surrounding the source.  The net count rate was
$\sim$0.149$\pm$0.003 counts s$^{-1}$.

A response matrix was constructed specific to the off-axis angle of
SN1978K.  The matrix was corrected for the time-dependent absorption
using the fitted functional form which depends upon a single
parameter, the time from launch; for this observation, the time from
launch was 1179 days \citep{Plu03}.

\section{Analysis: {\it Chandra} ACIS}

Given the sharp point spread function of the {\it Chandra} mirrors, we
must be concerned with possible event pileup even though the large
off-axis angle mitigates the effects of pileup considerably.  The
spectral analysis was undertaken using a pileup model \citep{Davis01}.
The resulting fit indicated low or zero pileup.  Given the broader
point spread function of the {\it XMM-Newton} mirrors, there is no
pileup of the EPIC-pn or MOS spectra from that instrument.  The
results of the fits to the {\it Chandra} and {\it XMM-Newton} spectra,
to be described below, are very similar, as one expects for
observations separated by $\sim$2 years of a slowly-evolving object.
The similarity supports the conclusion of low or zero pileup.  We
further fit the spectra with both XSPEC \citep{Arn96} and the Sherpa
fitting engine \citep{FDS01} and obtained identical results.

\begin{figure}
\begin{center}
\caption{Spectral fit to {\it Chandra} ACIS spectrum using a
Raymond-Smith model adopted from the best-fit {\it ASCA} model
\citep{Petre94}.  For this figure, only the model normalization was
adjusted.  Spectral evolution from the {\it ASCA} 1993 epoch is
evident.}
\label{oldfit}
\plotone{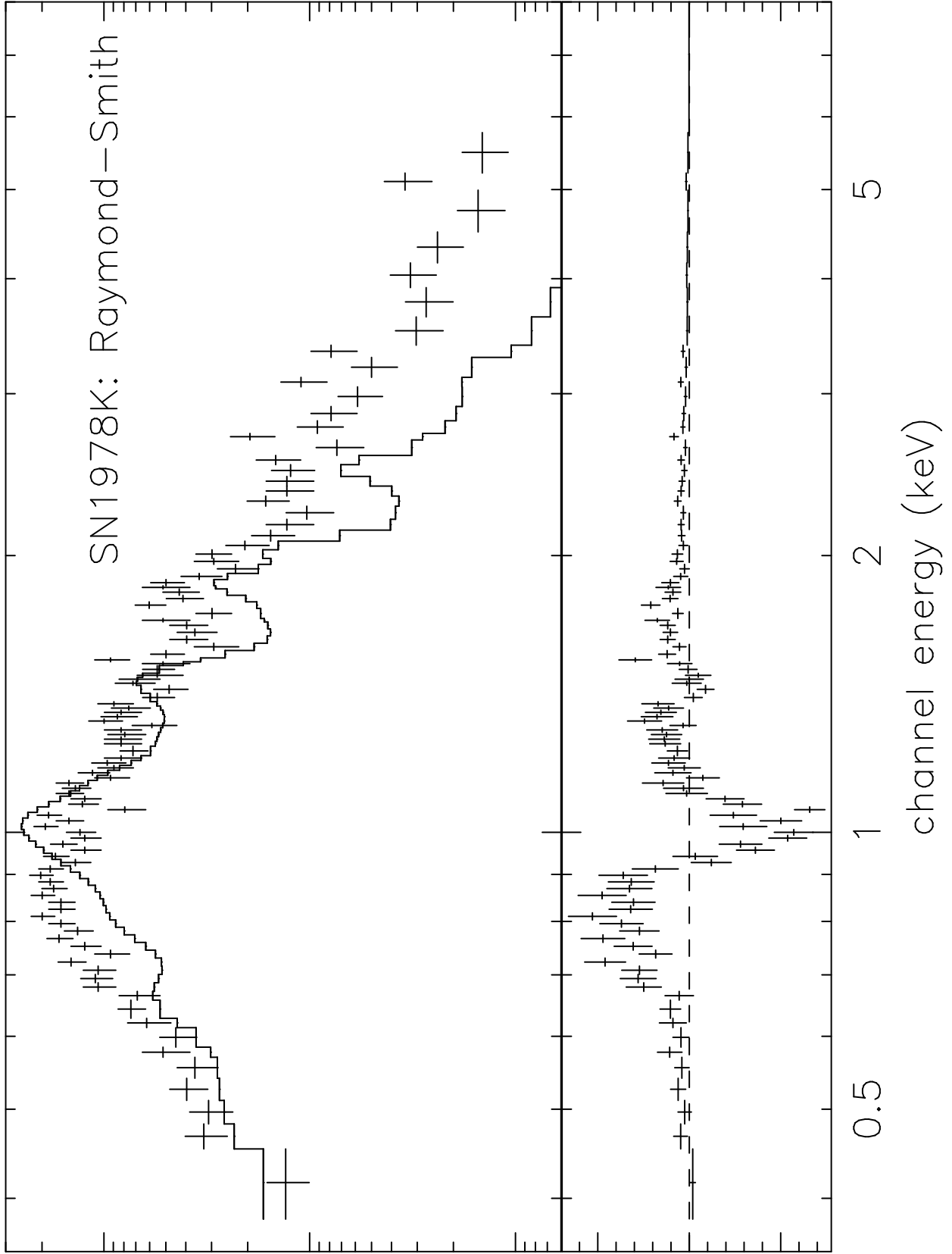}
\end{center}
\end{figure}

We first used the best-fit models derived from the {\it ROSAT} PSPC
and {\it ASCA} SIS/GIS observations of SN1978K which consisted of
absorbed, single-component continuum models (R93, \citealt{Petre94},
S99).  We fixed the model parameters at the previously-determined
values, but it was immediately evident that the models no longer
provided a good fit, instead yielding a ${\chi}^2/{\nu}$ $\geq$3.  We
then allowed the model parameters to vary; the resulting model fits
remained poor.  Figure~\ref{oldfit} shows the fit using a single
absorbed Raymond-Smith model.  Clearly a single-component model is
inadequate as it can not simultaneously fit the emission in the
0.7-1.2 keV region as well as the apparent hard component at energies
above $\sim$2-3 keV.  That the {\it ROSAT} PSPC models do not provide
a good fit is not especially surprising given the softer response of
that detector.  That the {\it ASCA} spectra do not provide a good fit
may be explained as either an increase in flux above 2 keV by a factor
of $\sim$4 since 1995 or by the difficulty in carrying out background
subtraction for the {\it ASCA} spectra stemming from the large
($\sim$1$'$.5) point spread function.  Indeed, the fit to the {\it
ASCA} spectra suggests the presence of a harder component that was not
statistically significant.  We believe a factor of 4 is too large to
be ascribed to difficulties with background subtraction.

We then explored a variety of models.  We do not include a long
discussion of those models that failed, but instead focus on the more
successful ones.  Most of the models failed by missing flux in the
0.8-1.0 keV band, a region known to contain the potential for
considerable line emission.  Table~\ref{specfit} includes a listing of
the models ranked by increasing ${\chi}^2/{\nu}$.  The `dual brems'
and `brems+multi-gauss' models yield adequate fits but leave
systematic residuals.  The non-equilibrium ionization model (in
`xspec' lingo, `nei'; \citealt{Bork01} and references therein), the
plane-parallel shock model (`pshock'; \citealt{Bork01}) as well as
other shock models also provided adequate fits, but did not achieve
the lowest ${\chi}^2$ values.  The ionization equilibrium collisional
plasma model (`equilib') yielded a very poor fit.  The single
component power law, bremsstrahlung, and Raymond-Smith models were
included for direct comparison with the historical results from {\it
ROSAT} and {\it ASCA} (e.g., S99, \citealt{Petre94}).

\begin{figure}
\begin{center}
\caption{Dual variable Mekal model fit to the {\it Chandra} ACIS
spectrum.  The dashed lines show the individual model components.
Note the strong Si emission in the model of the soft component.}
\plotone{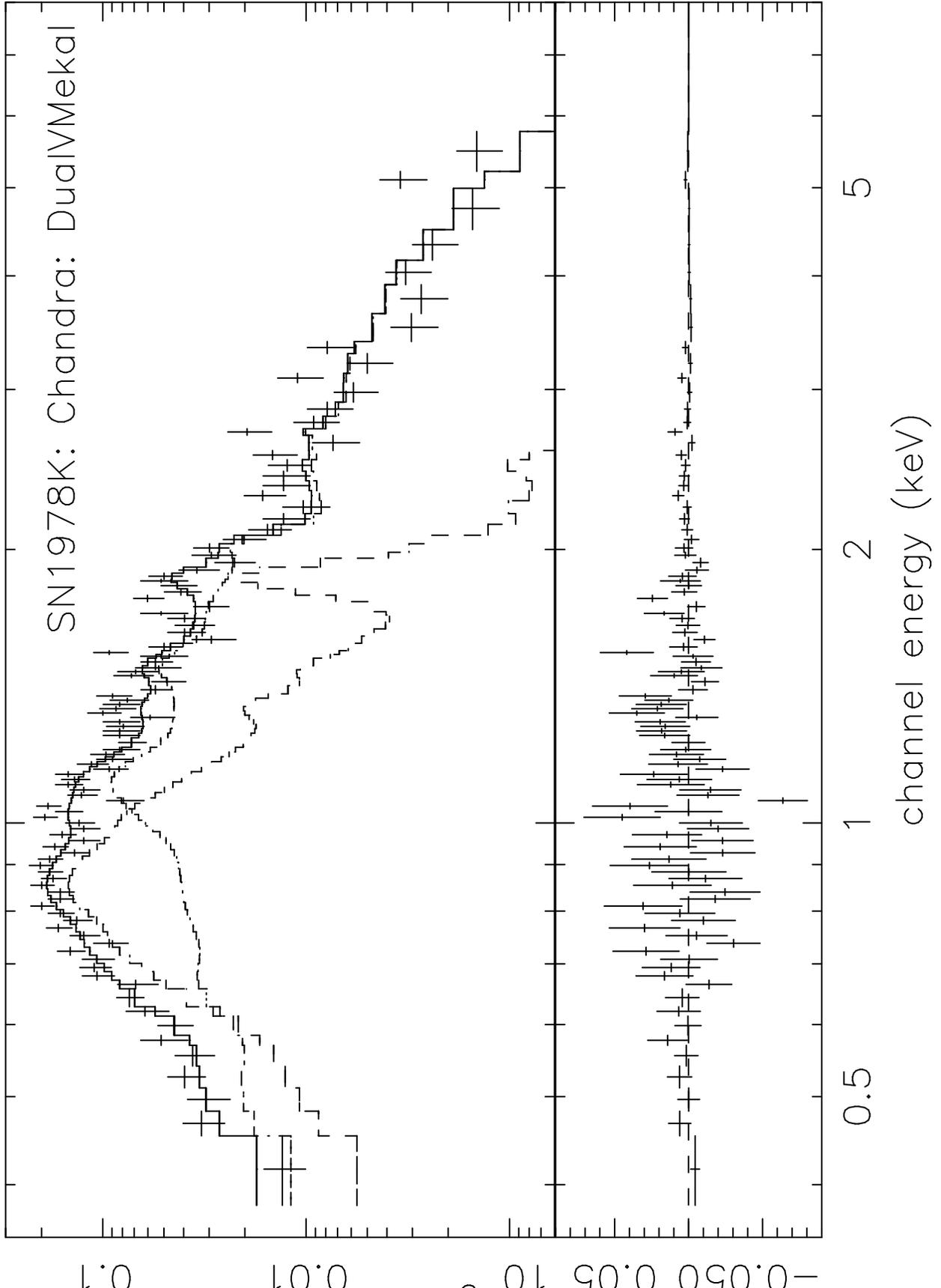}
\label{fig-dvmek}
\end{center}
\end{figure}

\begin{figure}
\begin{center}
\caption{Contour plot for the (temperature, column density) parameters
of the dual variable Mekal model for {\it Chandra} and {\it
XMM-Newton}.  The upper set of contours in each figure belong to {\it
Chandra}, the lower set to {\it XMM-Newton}. a) low-temperature
component and column density; (b) high-temperature component and
column density.}
\plotone{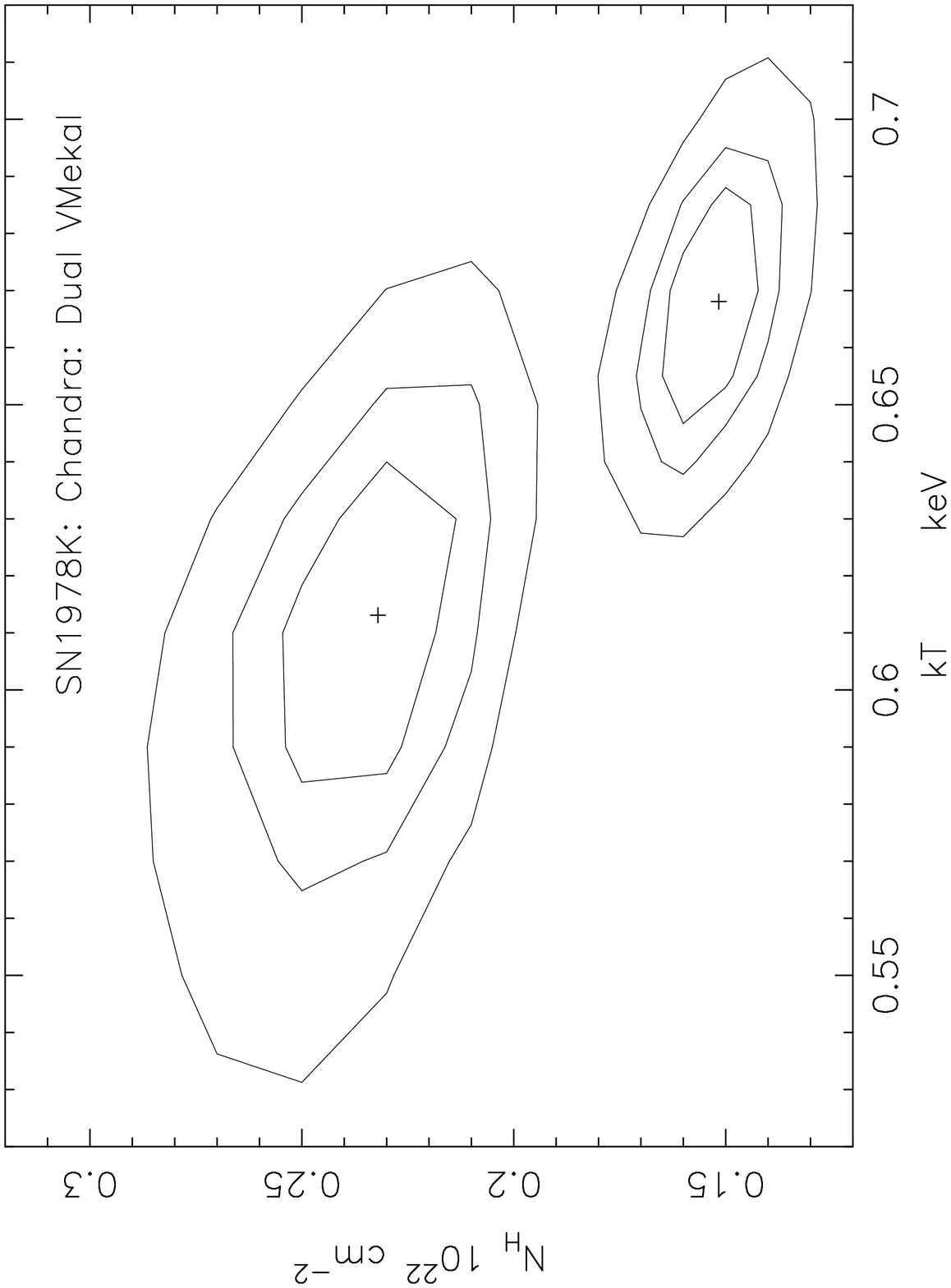}
\plotone{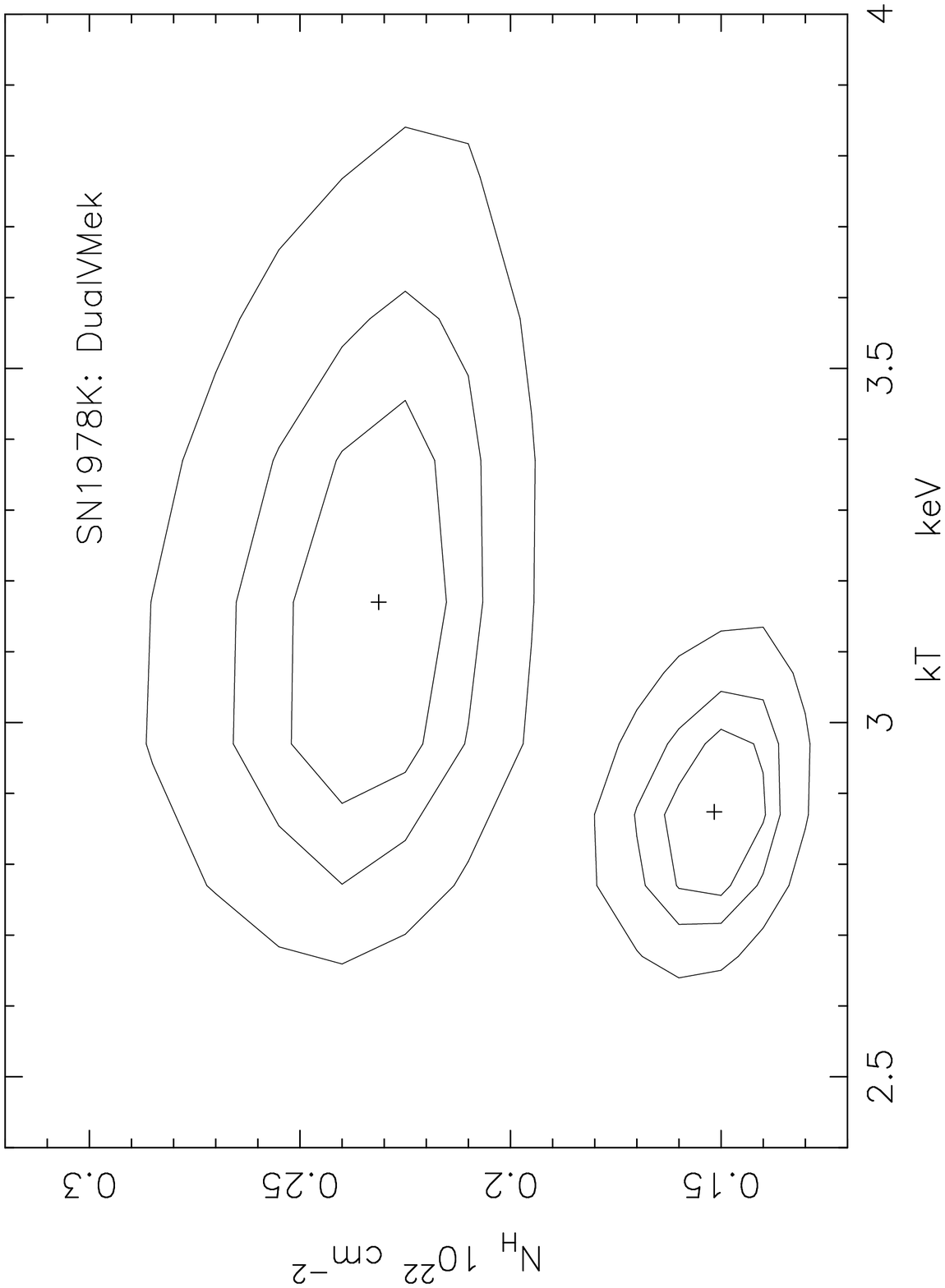}
\label{cont_dvmek}
\end{center}
\end{figure}

The best-fit model was an absorbed, two-temperature, optically thin
thermal plasma model (the variable Mekal model `vmekal' in xspec lingo
that uses the line calculations of \cite{Mewe1}, \cite{Mewe2}, and
\cite{Kaastra92} plus the Fe L enhancements of \cite{Lied95}).
Figure~\ref{fig-dvmek} shows the fitted spectrum.  The best-fit
temperatures are 0.61$^{+0.04}_{-0.05}$ and 3.16$^{+0.44}_{-0.42}$ keV
and are shown as the {\it upper} set of contour plots in
Figure~\ref{cont_dvmek} (the lower set of contours in each figure will
be discussed shortly).  The model-derived fluxes are listed in
Table~\ref{fluxval} and generally fall in the range of
4--7$\times$10$^{-13}$ erg s$^{-1}$ cm$^{-2}$ in the 0.5-2 keV band
and $\sim$3--4$\times$10$^{-13}$ erg s$^{-1}$ cm$^{-2}$ in the 2-10
keV band with the low numbers corresponding to the absorbed flux and
the high to the unabsorbed.  These fluxes correspond to a 0.5-2 keV
luminosity of $\sim$7.5--14$\times$10$^{38}$ erg s$^{-1}$ and a 2-10 keV
luminosity of $\sim$6--7.5$\times$10$^{38}$ erg s$^{-1}$.

The best-fit column, $\sim$2.3$\times$10$^{21}$ cm$^{-2}$ and shown in
Figure~\ref{cont_dvmek}, lies about a factor of 8--10 above the
measured column in the direction of NGC 1313,
$\sim$3.7$\times$10$^{20}$ cm$^{-2}$ \citep{SFD98}.  However, it does
correspond to within 30\% with the E$_{\rm B-V}$ value of 0.31
measured from optical spectra \citep{Ryder93} and using the N$_{\rm
H}$-E$_{\rm B-V}$ conversion of \cite{PS95}: N$_{\rm H}$
$\sim$5.3$\times$10$^{21}$ E$_{\rm B-V}$.

\begin{figure}
\begin{center}
\caption{Contour plot for Si abundances for the dual variable Mekal
model.  Solid lines are the {\it Chandra} contours; dashed lines are
the {\it XMM-Newton} contour.  The Si abundances of the low-T
components differ from solar at the 90\% level (middle contour rings)
while those of the high-T components are essentially undetermined.}
\plotone{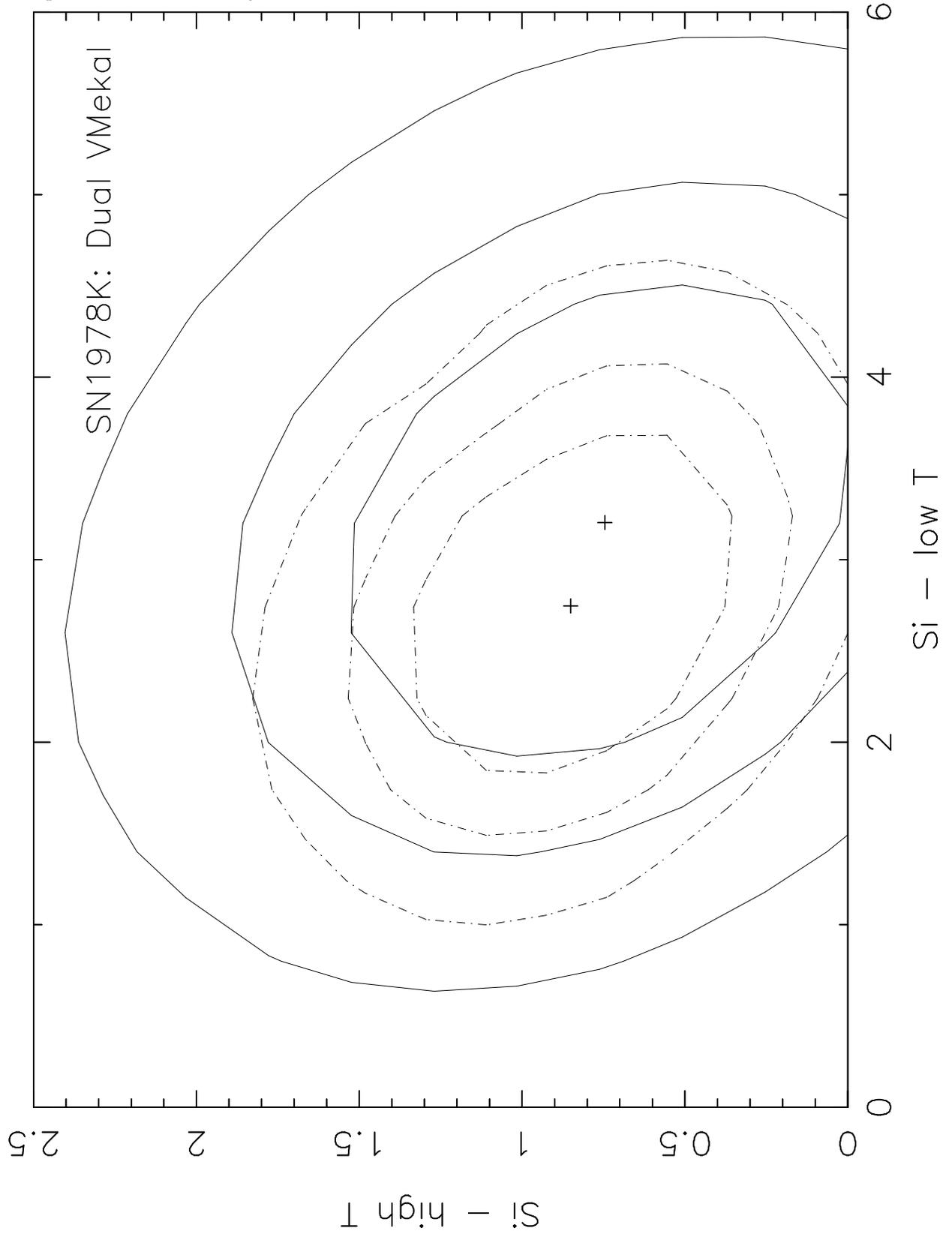}
\label{cont_si}
\end{center}
\end{figure}

The vmekal model permits varying the abundances of
astrophysically-important elements.  We varied the abundances of each
element in turn but forced the corresponding abundances of the soft
and hard components to vary simultaneously but independently.  This
approach ensured the most robust detection of specific line features
as well as differences between the soft and hard components.  The only
abundance inconsistent with solar is that of Si for the soft
component.  Figure~\ref{cont_si} shows the uncertainty contours for
the abundance of Si.  The soft component abundance is 3.20 which is
significantly different from solar at the 90\% confidence level.  The
Si abundance for the hard component is consistent with solar.  The
fluxes for each component are listed separately in
Table~\ref{fluxval}.

\section{Analysis: {\it XMM-Newton}, {\it ASCA}, \& {\it ROSAT}}

We also extracted spectra from the {\it XMM-Newton} observation of
2000 October 17.  The observation lasted for 42416 sec but high
background significantly constrained the extracted spectra to shorter
times.  The EPIC pn spectrum totaled 15940 s with an average rate of
0.263 counts s$^{-1}$; the MOS-1 and MOS-2 spectra totaled 23838 and
24124 sec, respectively, at an average count rate of 0.088 and 0.089
counts s$^{-1}$.  Just under 50\% of the total extracted counts formed
the pn spectrum.

\begin{figure}
\begin{center}
\caption{Plot of the {\it XMM-Newton} spectrum (EPIC pn, MOS-1, and -2) with the fitted dual
variable Mekal model.}
\plotone{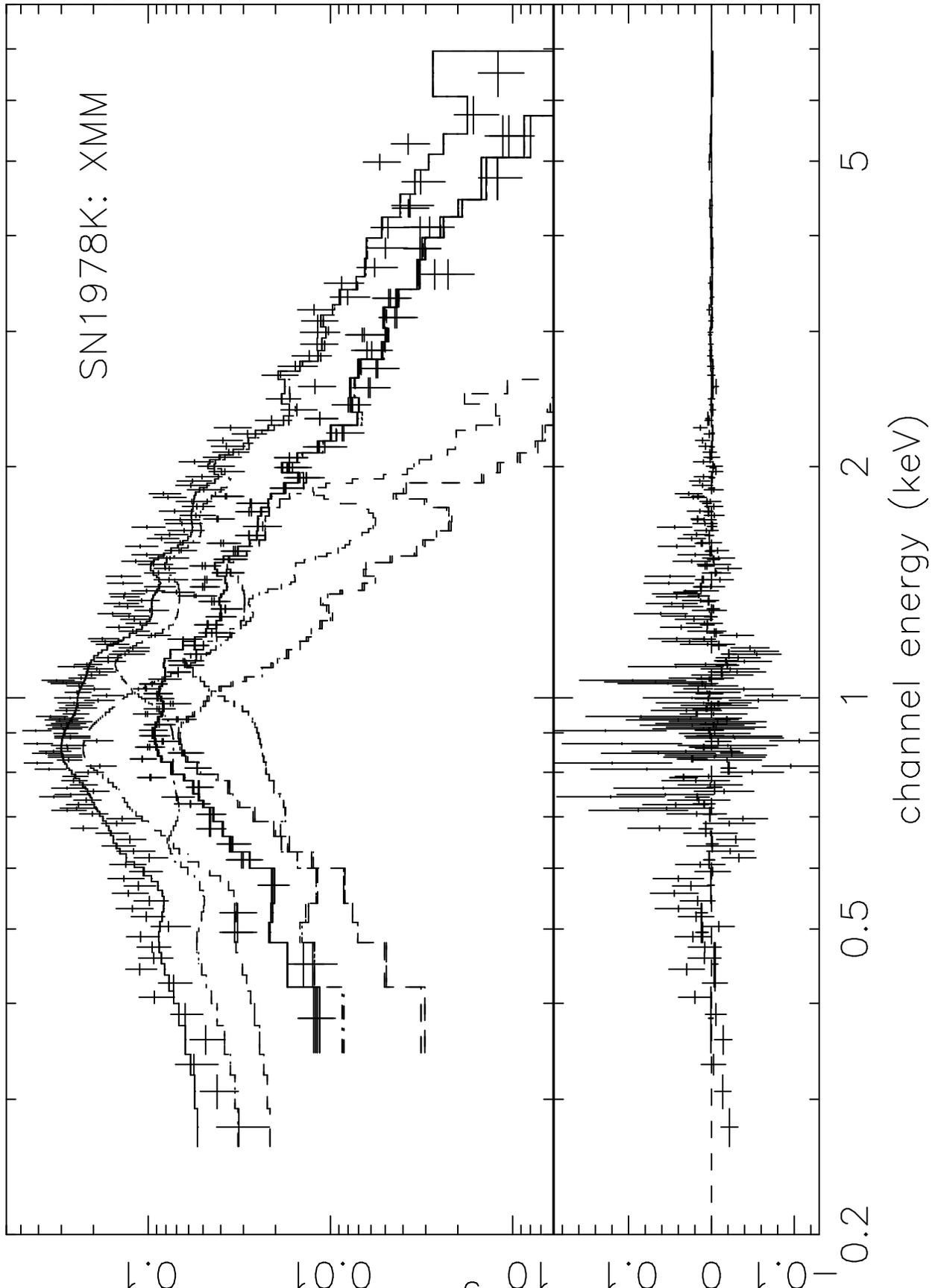}
\label{fit_xmm}
\end{center}
\end{figure}

We fit the pn and MOS spectra using the best-fit {\it Chandra} model,
namely the dual variable Mekal model.  Table~\ref{ascafit} lists the
resulting model parameters.  The fitted spectrum is shown in
Figure~\ref{fit_xmm}.  The best-fit temperatures are 0.67 and 2.87
keV.  Rather than plot the parameter contours separately, we included
them on the {\it Chandra} ACIS plots of Figure~\ref{cont_dvmek}; the
{\it XMM-Newton} contours are the {\it lower} set of contours in each
figure.  Note that the temperatures measured from the two instruments
are essentially identical within the errors; the contours differ only
in N$_{\rm H}$ by a factor of $\sim$1.65.  One possible explanation is
an increasing column density as the material cools and recombines.

The inferred luminosity in the soft band depends very sensitively on
the adopted value for N$_{\rm H}$; because of that sensitivity, we
adopt for N$_{\rm H}$ a value determined from the measured E$_{\rm
B-V}$ for the {\it ROSAT} HRI and PSPC spectra as well as the {\it
ASCA} spectra.  This approach assigns a reasonable value to N$_{\rm
H}$ for the HRI data points for which no spectral information would
otherwise be available.  In addition, adoption of an E$_{\rm
B-V}$-determined value for N$_{\rm H}$ removes the artificially high
value determined from fits to the {\it ASCA} spectra for which the
calibration below $\sim$0.7 keV is uncertain \citep{SCU00}.  We use
the {\it Chandra}- and {\it XMM-Newton}-determined values of N$_{\rm
H}$ to probe the sensitivity of the unabsorbed fluxes to the adopted
values.

For direct comparison of the {\it Chandra} and {\it Newton} results
with the previously-described {\it ASCA} spectra (\citealt{Petre94},
S99), we re-fit the {\it ASCA} spectra using the dual variable Mekal
model as well as a single variable Mekal model.  We included a single
variable Mekal model to test whether the dual model is statistically
required to fit the SIS spectra.  Necessarily, the lower count rate
yields larger error bars for each parameter of the dual Mekal model.
We restrict our attention solely to the SIS spectra to ensure the best
sensitivity to line emission.  The fitted parameters values are listed
in Table~\ref{ascafit} while the fluxes are listed in
Table~\ref{fluxval}.  From the ${\chi}^2/{\nu}$ values, the dual
variable model provides a statistically better fit.

For the {\it ROSAT} PSPC and HRI fluxes, we re-extracted and re-fit
the PSPC spectra using a single variable Mekal model.  A dual model is
not statistically required to provide a good fit largely because of
the narrow bandpass.  All of the abundance values were consistent with
solar, within the errors, so we froze all at solar values.  Unabsorbed
fluxes were calculated in the 0.5-2 keV band.  As HRI does not possess
spectral resolution, we must adopt a fixed spectral model.  We used a
`best fit' model of a single variable Mekal spectrum at kT = 0.6 keV
and N$_{\rm H}$ = 1.8$\times$10$^{21}$ cm$^{-2}$.  This spectrum is
essentially an average of the PSPC spectra and adopting the E$_{\rm
B-V}$-determined N$_{\rm H}$ value.  Only the model normalization was
adjusted for each HRI point.  Unabsorbed fluxes in the 0.5-2 keV band
were then calculated.  As a check, we converted count rates using the
PIMMS\footnote{PIMMS=Portable, Interactive Multi-Mission Software;
\cite{Mukai93}.} code; all of the converted values were within the
errors of the calculated unabsorbed fluxes.  We expect the fluxes are
accurate within 25-30\%.

\section{Light Curves}

From the collected set of unabsorbed fluxes, we constructed a light
curve of the X-ray emission of SN1978K including an upper limit from
the {\it Einstein} IPC reported in \cite{Ryder93,Schl96} in the soft
(0.5-2 keV) and hard (2-10 keV) bands.

\begin{figure}
\begin{center}
\caption{X-ray light curve in the 0.5-2 keV band based upon
observations obtained with the {\it ROSAT} PSPC, {\it ROSAT} HRI, {\it
ASCA} SIS/GIS, {\it XMM-Newton} EPIC, and {\it Chandra} ACIS and
including an upper limit from an {\it Einstein} IPC observation (lower
left corner).}
\plotone{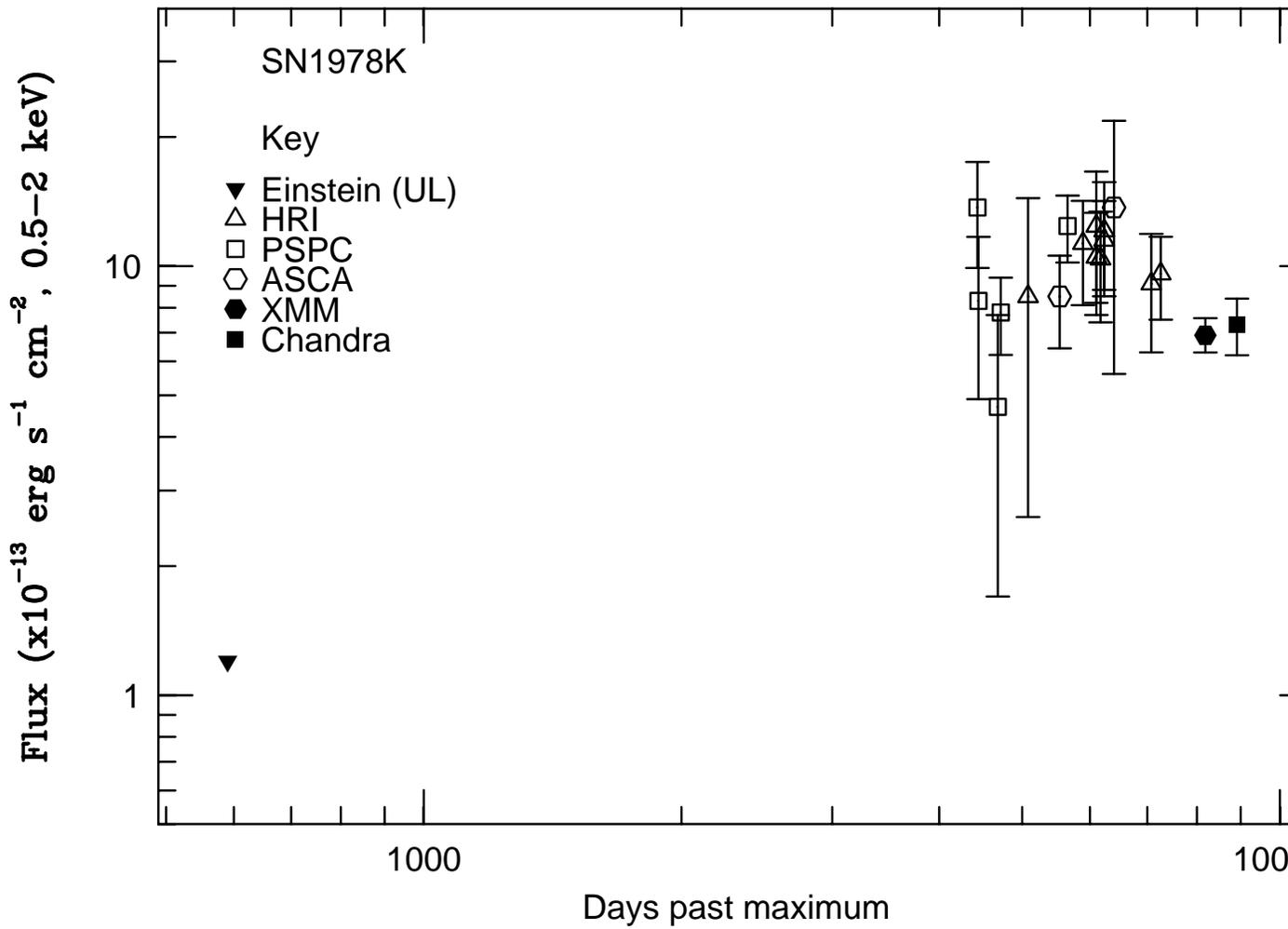}
\label{full_lc}
\end{center}
\end{figure}

The soft band light curve (Figure~\ref{full_lc}) shows immediately
that the behavior in the 0.5-2 keV band is completely determined by
the adopted value for the column density.  Formally, we see a
$\sim$20\% decline from the {\it ASCA} epochs but within the errors,
the light curve may still be interpreted as constant.  

\begin{figure}
\begin{center}
\caption{X-ray light curve in the 2-10 keV band based upon
observations obtained with {\it ASCA} SIS/GIS, {\it XMM-Newton} EPIC,
and {\it Chandra} ACIS and including an upper limit from an {\it
Einstein} IPC observation (lower left corner).  The IPC limit is
extrapolated from the 2-4.5 keV band of the detector to the 2-10 keV
band assuming a 3 keV bremsstrahlung model.}
\plotone{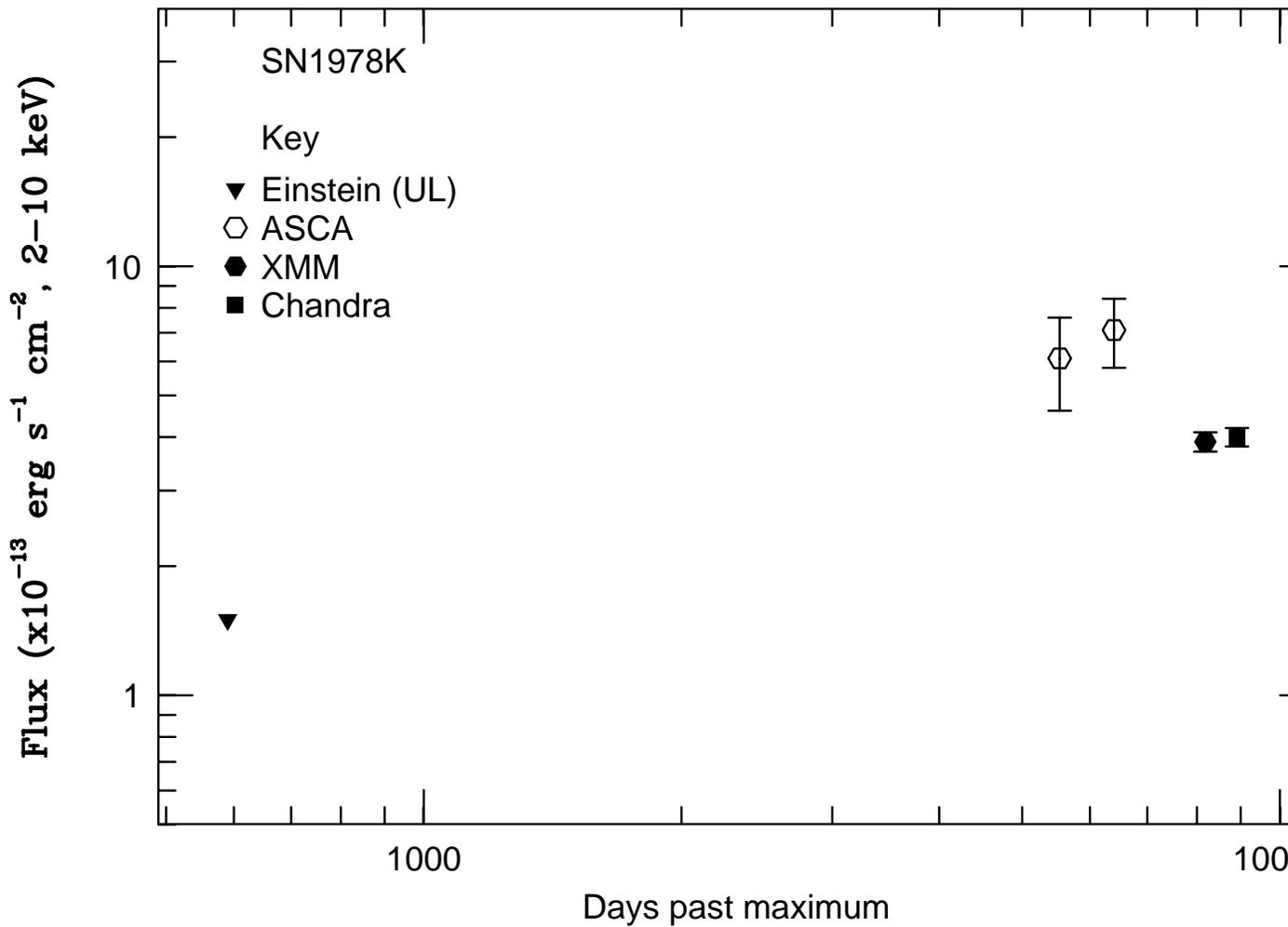}
\label{hard_lc}
\end{center}
\end{figure}

The hard band light curve (Figure~\ref{hard_lc}) shows a decline of
about a factor of 1.5 (mean {\it ASCA} values
$\sim$6.5$\times$10$^{-13}$ erg s$^{-1}$ cm$^{-2}$ vs. {\it XMM}/{\it
Chandra} values of $\sim$4$\times$10$^{-13}$ erg s$^{-1}$ cm$^{-2}$).
If the factor 1.5 decline is indicative of a future trend, then the
peak occurred approximately 6000 days post-maximum, or near October
1994.  For both the soft and hard light curves, the `peak' is broad.

The {\it XMM-Newton} and {\it Chandra} points may be samples from a
`hesitation' on the decline.  Such behavior, while completely
speculative without additional observations, nevertheless is not
without precedent - SN1979C \citep{Montes00} and SN2001ig
\citep{Ryd04} both showed hesitations or outright increases during the
decline phase of the radio light curves.  The corresponding X-ray
behavior is wholly unknown.

The behavior of the X-ray light curves is in contrast to the radio
light curves all of which peaked considerably earlier.  The fitted
times of radio maximum differ because the radio light curve behavior
is a function of frequency.  However, the band with the latest peak, L
at 1.38 GHz, reached a maximum near age 2000 days, or approximately
November 1983 (S99), or about 11 years earlier.

\section{Discussion}

The apparent decline in the hard band vs. the soft band coupled with
the potential emergence of a Si emission line signal the increasing
transparency of the circumstellar medium surrounding SN1978K.  If the
X-ray transparency is increasing, then SN1978K has started its decline
toward its remnant stage.  Future observations will reveal the trend
more concretely.  We expect increasing line emission in the coming
years.

However, we must be more cautious as the possibility exists that the
column density is increasing as the material behind the shock
recombines.  A conclusion that the column density is increasing is at
this point premature, resting as it would on two data points (the {\it
Chandra} and {\it XMM} spectra).  The {\it ASCA} spectra
insufficiently constrained the column density to provide measurements
of the trend over a longer time base.  A future observation with {\it
Chandra} or {\it Newton} would, with similar statistics per spectral
channel, establish the nature of any trend that may exist.

The power law indices (n, s) of the density profile of the ejecta
(${\rho}_{\rm ejecta}$ $\propto$~r$^{\rm -n}$) and circumstellar matter
distribution (${\rho}_{\rm cs}$ $\propto$~r$^{\rm -s}$) are important
parameters of interaction models.  They may be calculated from the
light curve and from the ratio of the temperatures of the forward and
reverse shocks (e.g., \citealt{Fran96}).

We interpret the temperatures of the two components as the
temperatures of the forward and reverse shock emission.  \cite{Fran96}
show that the temperature ratio depends upon the circumstellar and
supernova ejecta density distributions as ${T_{\rm r}}/{T_{\rm f}} =
(3 - s)^2 / ( n - 3 )^2,$ where s = circumstellar density distribution
exponent, usually 2 for a constant and homogeneous stellar wind, and n
= SN ejecta density distribution exponent.  The ratio of the
temperatures of the two model fits to the {\it Chandra} spectrum is
0.61/3.2 = 0.19; solving for n yields an index of 5.25$\pm$0.25.
Repeating the same arithmetic for the {\it XMM-Newton} spectrum yields
an index of 5.07$\pm$0.10.  An index of $\sim$5 is lower than usually
adopted for the ejecta distribution; for example, \cite{CF94} derive
results covering the index range of 7-12.  

We argue, however, that n $\sim$5 is reasonable.  Recently,
\cite{Zim03} described an {\it XMM-Newton} observation of SN1993J in
M81 and obtained a best-fit spectrum using the dual variable Mekal
model.  They inferred n = 8.9 from the fitted temperatures after
fitting the light curve to obtain s $\sim$ 1.65.

However, \cite{Fran98} showed that for SN1993J the circumstellar
matter density was not s $\sim$ 1.65 as they first estimated from the
radio light curve \citep{Fran96}, but was s = 2 because their initial
analysis failed to include synchrotron self-absorption.  Their
correction implies that circumstellar matter distributions really are
deposited with s = 2.  We therefore recalculate the index value for
SN1993J.

From their best-fit model, \cite{Zim03} obtained temperatures of
0.34$\pm$0.04 and 6.54$\pm$4 keV for the {\it XMM-Newton} spectrum.
Adopting s = 2, we obtain a value of n of $\sim$7.4$^{+1.5}_{-1.8}$,
where the large error range, driven by the large error on the
high-temperature component, just overlaps the s $\sim$1.65 value.  The
ejecta distribution index of SN1993J is slightly larger than the value
for SN1978K, but the error range reaches nearly to the SN1978K value
of n = 5.  

Having established that an index of $\sim$5 is not substantially
different from the the index of SN1993J, we briefly examine whether
the difference is in fact sensible.  From \cite{CF94}, lower values
for the index lead to smaller density contrasts between the forward
and reverse shocks, and adiabatic shocks at early times.  SN1978K was
essentially constant for at least 3 years near the peak (from
$\sim$day 5500 to day $\sim$6500; S99) while SN1993J declined
essentially from the time of the explosion \cite{Swartz03}.  The
increase in flux of SN1978K from the {\it Einstein} upper limit
suggests that absorption was critical at very early times.  This
behavior is consistent with the adopted model for Type IIn supernovae,
namely, an object with a dense circumstellar medium resulting from a
dense wind.  Furthermore, \cite{Fran96} showed that the cooling time
increases for smaller values of n.  The different behavior of SN1993J
(fading from start) and SN1978K (rises, then constant) fit within the
context of the circumstellar model.  The surprise is the apparent
extreme sensitivity across a small difference in n.

We caution the reader to consider these numbers skeptically.  As
pointed out by, for example, \cite{Swartz03}, extrapolation of the
self-similar interaction model to phases beyond a few months to years
is ``inappropriate'' to adopt their description.  In particular, the
forward shock rapidly moves into prior phases of mass loss from the
progenitor which need not be distributed in a smooth $r^{-2}$ profile
as usually assumed.  Numerical calculations become necessary,
including radiative losses, particularly for the reverse shock.  We
note the presentation of \cite{NF03} demonstrating that multiple
temperatures exist in the shocked gas in contrast to the considerably
simpler models adopted for the present analysis.

\section{Conclusions}

We conclude that the {\it Chandra} and {\it XMM-Newton} observations
detected the first hints of increasing transparency of the matter
surrounding SN1978K.  The increasing transparency signals the start of
its decline in the X-ray band.

The low-temperature spectral components remain essentially unchanged
from the {\it ASCA} epochs except for the addition of the detected Si.
Within the errors, the high-temperature components have also not
changed; if we restrict our attention solely to the {\it XMM-Newton}
and {\it Chandra} observations of 2000 and 2002, then we see a decline
of a factor of $\sim$1.5, consistent with an interpretation of
increasing transparency.  Additional X-ray observations will confirm
or refute our interpretation.

If X-ray line emission is becoming increasingly visible, line emission
in other bands may be expected to change depending upon the atomic
cascades.  An observation with the {\it HST} STI Spectrograph is
clearly warranted; such an observation would also aid in sorting out
the emission region and clarifying the nature of the line emission
uncovered with the FOS observation (S99).

Additional X-ray observations over the coming years are important to
investigate possible variations in the wind as well as detecting the
expected developing line emission.  If our interpretation is correct,
then the decline phase commenced just within the past few years and
provides an opportunity to observe the decline.  Comparison with
SN1987A will reveal differences likely attributable to the differing
densities of the circumstellar media.  

At least two radio-emitting supernovae have shown hesitations and
increases on the decline, interpreted as enhanced mass loss during the
late stages of stellar evolution prior to the progenitor's explosion
(e.g., SN2001ig, \cite{Ryd04}; SN1993J, \cite{Zim03}; SN1979C,
\cite{Montes00} and references therein).  With the shock moving at
about 10$^3$ times the wind velocity, continued X-ray and radio
observations are expected to provide in a human lifetime an extended
view of several thousand years of late wind and mass loss behavior of
a massive star.

\acknowledgements

The research of EMS was supported by contract number NAS8-39073 to
SAO. PK acknowledges partial support from NASA grant NAG5-7405 and
{\it Chandra} grant GO2-3102X.

\begin{table*}
\begin{center}
\caption{Spectral Model Fits to ACIS Spectrum\tablenotemark{a}}
\label{specfit}
\begin{tabular}{rrrrrrrr}
 Model & ${\chi}^2/{\nu}$ & DoF & N$_{\rm H}$ (x10$^{22}$ cm$^{-2}$) & Param-1 & Param-2 & Norm-1 & Norm-2 \cr \hline
Dual VMekal & 1.20 & 100 & 0.23$^{+0.04}_{-0.03}$ & T$_1$=0.61$^{+0.04}_{-0.05}$ & T$_2$=3.16$^{+0.44}_{-0.40}$ & 1.4e-4 & 4.9e-4 \cr
  + vary Si & $\cdots$ & $\cdots$ & $\cdots$ & Si=3.20$^{+1.80}_{-1.90}$ & Si=0.70$^{+1.20}_{-0.70}$ & $\cdots$ & $\cdots$ \cr
\cr
Dual Brems & 1.62 & 99 & 0.47$^{+0.30}_{-0.02}$ & T$_1$=0.27$^{+0.01}_{-0.10}$ & T$_2$=3.42$^{+1.90}_{-2.10}$ & 2.8e-4 & 6.6e-2 \cr
~~+ Gauss\tablenotemark{b} & $\cdots$ & $\cdots$ & $\cdots$ & E$_{\rm line}$=0.877$^{+0.03}_{-0.10}$ & EqW=73$^{+79}_{-46}$ & $\cdots$ & $\cdots$ \cr
~~+ Gauss\tablenotemark{b} & $\cdots$ & $\cdots$ & $\cdots$ & E$_{\rm line}$=1.80$^{+0.07}_{-0.02}$ & EqW=78$^{+24}_{-44}$ & $\cdots$ & $\cdots$ \cr
\cr
Brems +  & 1.67 & ~98 & 0.17$^{+0.04}_{-0.03}$ & T=1.89$^{+0.41}_{-0.23}$ & $\cdots$ & 3.8e-4 & $\cdots$ \cr
~~+ Gauss\tablenotemark{b,c} & $\cdots$ & $\cdots$ & $\cdots$ & E$_{\rm line}$=0.654 & EqW=79$^{+78}_{-38}$ & $\cdots$ & $\cdots$ \cr
~~+ Gauss\tablenotemark{b,c} & $\cdots$ & $\cdots$ & $\cdots$ & E$_{\rm line}$=0.826 & EqW=165$^{+55}_{-46}$ & $\cdots$ & $\cdots$ \cr
~~+ Gauss\tablenotemark{b,c} & $\cdots$ & $\cdots$ & $\cdots$ & E$_{\rm line}$=0.915 & EqW=101$^{+30}_{-41}$ & $\cdots$ & $\cdots$ \cr
~~+ Gauss\tablenotemark{b,c} & $\cdots$ & $\cdots$ & $\cdots$ & E$_{\rm line}$=1.02 & EqW=122$^{+29}_{-54}$ & $\cdots$ & $\cdots$ \cr
~~+ Gauss\tablenotemark{b,c} & $\cdots$ & $\cdots$ & $\cdots$ & E$_{\rm line}$=1.86 & EqW=60$^{+29}_{-54}$ & $\cdots$ & $\cdots$ \cr
\cr
P-shock & 1.87 & 104 & 0.69$^{+0.03}_{-0.03}$ & T=2.29$^{+0.60}_{-0.43}$ & log ${\tau}$=9.954$^{+0.059}_{-0.051}$ & 9.1e-4 & $\cdots$ \cr
NEI & 2.06 & 104 & 0.72$^{+0.03}_{-0.03}$ & T=2.42$^{+0.58}_{-0.42}$ & log ${\tau}$=9.714$^{+0.038}_{-0.035}$ & 8.4e-4 & $\cdots$ \cr 
Powerlaw & 3.12 & 105 & 0.33$^{+0.05}_{-0.03}$ & ${\Gamma}$=3.12$^{+0.24}_{-0.17}$ & $\cdots$ & 5.5e-4 & $\cdots$ \cr
Brems & 3.57 & 105 & 0.18$^{+0.02}_{-0.04}$ & T=1.25$^{+0.32}_{-0.12}$ & $\cdots$ & 6.2e-4 & $\cdots$ \cr 
Equilib\tablenotemark{d} & 4.59 & 105 & 0.06 & kT=2.74 & $\cdots$ & 5.9e-4 & $\cdots$ \cr
Raymond-Smith & 6.86 & 105 & 0.08$^{+0.04}_{-0.03}$ & T=1.09$^{+0.02}_{-0.02}$ & $\cdots$ & 3.3e-4 & $\cdots$ \cr \hline
\end{tabular}
\end{center}

\tablenotetext{a}{All errors listed are 90\%.  Units are 10$^{22}$ cm$^{-2}$ for N$_{\rm H}$, keV for kT; eV for line energy and line widths.}
\tablenotetext{b}{Gaussian line width fixed at 0.0 to model an
unresolved line; the line is broadened by the intrinsic energy
resolution of the ACIS detector.}
\tablenotetext{c}{Gaussian line energy fixed at listed value.}
\tablenotetext{d}{No error bars are included on ``Equilib'' because
the fit was too poor.  Error bars would not normally be included for
the others because of the poor fits but are so included for comparison
with the historical {\it ROSAT} and {\it ASCA} results.}

\end{table*}

\begin{table*}
\begin{center}
\caption{Fluxes for Best-Fit Models: {\it Chandra}, {\it XMM-Newton}, and {\it ASCA}\tablenotemark{a}}
\label{fluxval}
\begin{tabular}{lllllllll}
       & Age\tablenotemark{b} &  & \multicolumn{2}{c}{Complete Model} & \multicolumn{2}{c}{Soft Component} & \multicolumn{2}{c}{Hard Component} \cr
Satellite & (days) & Band  & Absorbed  & Unabsorbed & Absorbed  & Unabsorbed  & Absorbed  & Unabsorbed \cr \hline
Chandra & 8910 & 0.5-2 & 3.80$\pm$0.57 & 7.26$\pm$1.09 & 1.66$\pm$0.25 & 3.59$\pm$0.54 & 2.13$\pm$0.32 & 3.66$\pm$0.55 \cr 
        & & 2-10  & 3.87$\pm$0.23 & 3.98$\pm$0.24 & 0.10$\pm$0.01 & 0.11$\pm$0.01 & 3.81$\pm$0.23 & 3.92$\pm$0.24 \cr 
XMM     & 8184 & 0.5-2 & 4.61$\pm$0.41 & 6.86$\pm$0.62 & 1.87$\pm$0.17 & 2.98$\pm$0.27 & 2.72$\pm$0.24 & 3.88$\pm$0.35 \cr
        & & 2-10  & 3.92$\pm$0.01 & 3.95$\pm$0.16 & 0.11$\pm$0.01 & 0.12$\pm$0.01 & 3.76$\pm$0.18 & 3.83$\pm$0.16 \cr  
ASCA-1\tablenotemark{c}  & 5531 & 0.5-2 & 4.86$\pm$1.38 & 8.49$\pm$2.07 & 2.25$\pm$0.72 & 4.30$\pm$1.39 & 2.61$\pm$0.64 & 4.19$\pm$1.03 \cr
        & & 2-10  & 5.91$\pm$1.67 & 6.05$\pm$1.47 & 0.14$\pm$0.04 & 0.15$\pm$0.05 & 5.77$\pm$1.41 & 5.90$\pm$1.45 \cr
ASCA-2\tablenotemark{c}  & 6401 & 0.5-2 & 5.38$\pm$3.16 & 13.7$\pm$8.07 & 3.13$\pm$1.84 & 6.31$\pm$3.72 & 3.11$\pm$1.83 & 4.98$\pm$2.93 \cr
        & & 2-10  & 5.02$\pm$0.91 & 7.09$\pm$1.28 & 0.13$\pm$0.02 & 0.14$\pm$0.02 & 6.78$\pm$1.22 & 6.89$\pm$1.25 \cr \hline
\end{tabular}
\tablenotetext{a}{All fluxes have units of 10$^{-13}$ erg s$^{-1}$ cm$^{-2}$.}
\tablenotetext{b}{Age in days past maximum, adopted to be 1978 May 22, MJD 43650.}
\tablenotetext{c}{N$_{\rm H}$ value of 2$\times$10$^{22}$ falls within fitted range and adopted for flux calculations.}
\end{center}
\end{table*}

\begin{table*}
\begin{center}
\caption{Model Fits to {\it XMM-Newton} and {\it ASCA} Spectra}
\label{ascafit}
\begin{tabular}{lllllllllll}
     &  &                &     & N$_{\rm H}$ &  &  \cr
Obs  & Date  &${\chi}^2/{\nu}$  & DoF & (10$^{22}$ cm$^{-2}$) & T$_{\rm soft}$ & T$_{\rm hard}$ & Norm-1\tablenotemark{a} & Norm-2\tablenotemark{a} \cr  \hline
\multicolumn{9}{c}{Dual VMekal model}\cr
XMM   & 2000 Oct 17 & 1.19 & 308 & 0.15$^{+0.02}_{-0.01}$ & 0.67$^{+0.03}_{-0.03}$ & 2.87$^{+0.18}_{-0.12}$ & 1.2e-4 & 5.2e-4 \cr
     &  & $\cdots$ & $\cdots$ & $\cdots$ & Si=2.77$^{+1.30}_{-1.27}$ & Si=0.85$^{+0.65}_{-0.65}$ & $\cdots$ & $\cdots$ \cr
ASCA-1\tablenotemark{b} & 1993 Jul 12 & 0.76  & ~37 & $<$0.60 & 0.73$^{+0.27}_{-0.33}$ & 4.42$^{+10}_{-1.60}$ & 1.2e-4 & 6.6e-4 \cr
ASCA-2\tablenotemark{b} & 1995 Nov 29 & 1.02 & ~60 & 0.36$^{+0.40}_{-0.17}$ & 0.60$^{+0.19}_{-0.21}$ & 4.01$^{+7.0}_{-1.20}$ & 2.9e-4 & 7.9e-4\cr 
\cr
\multicolumn{9}{c}{Single VMekal model}\cr
ASCA-1\tablenotemark{b} & 1993 Jul 12 & 1.15  & ~42 & 0.14$^{+0.20}_{-0.12}$ & 1.06$^{+2.53}_{-0.50}$ & $\cdots$ & 9.2e-4 & $\cdots$ \cr
ASCA-2\tablenotemark{b} & 1995 Nov 29 & 1.20 & ~64 & 0.67$^{+0.20}_{-0.22}$ & 0.63$^{+0.07}_{-0.08}$ & $\cdots$ & 1.7e-3 & $\cdots$ \cr \hline
\end{tabular}
\tablenotetext{a}{Errors on normalizations are attached to the fluxes
as listed in Table~\ref{fluxval}.}
\tablenotetext{b}{Too few counts to fit Si abundance; value fixed at solar.}
\end{center}
\end{table*}

\begin{table*}
\begin{center}
\caption{Fluxes: {\it ROSAT} PSPC and HRI}
\label{pspcflux}
\begin{tabular}{lllll}
          & Approximate &  &         & \cr 
          & Observation &  &        & Flux\tablenotemark{c}\cr
 Detector &     Date\tablenotemark{a} & MJD & Age\tablenotemark{b} & $\times$10$^{-13}$ \cr \hline
 IPC  & 1980 Jan ~2\tablenotemark{d} & 44240 & ~590 &  $<$1.2 \cr
 PSPC & 1990 Jul 11\tablenotemark{e} & 48083 & 4433 & 13.7$\pm$3.8 \cr
 PSPC & 1990 Jul 23                  & 48095 & 4445 & ~8.3$\pm$3.4 \cr
 PSPC & 1991 Mar 18                  & 48333 & 4683 & ~4.7$\pm$3.0 \cr
 PSPC & 1991 Apr 24                  & 48370 & 4720 & ~7.8$\pm$1.6 \cr
 HRI  & 1992 May 6\tablenotemark{a}  & 48748 & 5080 & ~8.5$\pm$5.9 \cr
 PSPC & 1993 Nov ~3                  & 49294 & 5649 & 12.4$\pm$2.2 \cr
 HRI  & 1994 Jul~9\tablenotemark{a}  & 49542 & 5888 & 11.3$\pm$3.2 \cr
 HRI  & 1995 Feb ~5\tablenotemark{a} & 49753 & 6103 & 12.4$\pm$4.2 \cr
 HRI  & 1995 Feb ~6\tablenotemark{a} & 49754 & 6104 & 10.5$\pm$2.8 \cr
 HRI  & 1995 Apr 16\tablenotemark{a} & 49823 & 6173 & 10.4$\pm$3.0 \cr
 HRI  & 1995 Jun 15\tablenotemark{a} & 49883 & 6233 & 11.5$\pm$2.7 \cr
 HRI  & 1995 Jun 16\tablenotemark{a} & 49884 & 6234 & 12.1$\pm$3.6 \cr
 HRI  & 1997 Oct ~4\tablenotemark{a} & 50725 & 7076 & ~9.1$\pm$2.8 \cr
 HRI  & 1998 Apr ~5\tablenotemark{a} & 50908 & 7258 & ~9.6$\pm$2.1 \cr \hline
\end{tabular}
\tablenotetext{a}{MJD at center of observation when spanning multiple
days; those observations are footnoted.}
\tablenotetext{b}{units = days; based upon adopted date of maximum =
1978 May 22 = MJD 43650.}

\tablenotetext{c}{All fluxes in 0.5-2.0 keV band with units of erg
s$^{-1}$ cm$^{-2}$.  Adopted model is an absorbed vmekal, kT = 0.6
keV, N$_{\rm H}$ = 2.0$\times$10$^{21}$ cm$^{-2}$, based upon the
the estimated value of E$_{\rm B - V}$ of 0.31 from \cite{Ryder93}.}

\tablenotetext{d}{Observation obtained by {\it Einstein} IPC \citep{FTM84,FT87}.}

\tablenotetext{e}{Observation obtained during {\it ROSAT} All-Sky
Survey using PSPC-C. Sequence number = RS932904n00.  See S99 for a
listing of other sequence numbers.}

\end{center}
\end{table*}

\section{Figures}

\begin{thebibliography}

\bibitem[Chandra Proposers' Guide(2002)]{Chan02} Chandra Proposers'
Guide, Rev 5.0 (Cambridge, MA: Chandra X-ray Center)

\bibitem[Arnaud(1996)]{Arn96} Astronomical Data Analysis Software and
Systems V, A.S.P. Conference Series, Vol. 101, 1996, George H. Jacoby
\& Jeannette Barnes, eds., p. 17.

\bibitem[Borkowski, Lyerly, \& Reynolds(2001)]{Bork01} Borkowski, K.,
Lyerly, W. J., \& Reynolds, S. P. 2001, ApJ, 548, 820

\bibitem[Chevalier \& Fransson(1994)]{CF94} Chevalier, R. \& Fransson,
C. 1994, ApJ, 420, 268

\bibitem[Davis(2001)]{Davis01} Davis, J. E. 2001, ApJ, 562, 575


\bibitem[Dopita \& Ryder(1990)]{DopRyd90} Dopita, M. \& Ryder,
S. 1990, IAU Circular 4950

\bibitem[Fabbiano \& Trinchieri(1987)]{FT87} Fabbiano, G. \&
Trinchieri, G. 1987, ApJ, 315, 46

\bibitem[Fabbiano, Trinchieri, \& MacDonald(1984)]{FTM84} Fabbiano, G.,
Trinchieri, G., \& MacDonald, A. 1984, ApJ, 284, 65

\bibitem[Fransson \& Bj\"ornsson(1998)]{Fran98} Fransson, C. \&
Bj\"ornsson, C.-I. 1998, ApJ, 509, 861

\bibitem[Fransson et al.(1996)]{Fran96} Fransson, C., Lundqvist, P.,
\& Chevalier, R. 1996, ApJ, 461, 993

\bibitem[Freeman, Doe, Siemiginowska(2001)]{FDS01} Freeman, P., Doe,
S., \& Siemiginowska, A. 2001, SPIE, 4477, 76

\bibitem[Garmire et al.(2003)]{Gar03} Garmire, G. P., Bautz, Mark W.,
Ford, Peter G., Nousek, John A., \& Ricker, George R., Jr., SPIE,
4851, 28

\bibitem[Immler \& Lewin(2002)]{IL02} Immler, S. \& Lewin, W. 2002,
astro-ph/020223

\bibitem[Kaastra(1992)]{Kaastra92} Kaastra, J.S. 1992, An X-Ray
Spectral Code for Optically Thin Plasmas (Internal SRON-Leiden Report,
updated version 2.0)

\bibitem[Liedahl, Osterheld, \& Goldstein(1995)]{Lied95} Liedahl,
D. A., Osterheld, A. L., \& Goldstein, W. H. 1995, ApJ, 438, L115

\bibitem[M\'{e}ndez et al.(2002)]{Mendez02} M\'{e}ndez, B., Davis, M.,
Moustakas, J., Newman, J., Madore, B. F., \& Freedman, W. L. 2002, AJ,
124, 213

\bibitem[Mewe, Gronenschild, \& van den Oord(1985)]{Mewe1} Mewe, R.,
Gronenschild, E.H.B.M., and van den Oord, G.H.J. 1985, A\&AS, 62, 197

\bibitem[Mewe, Lemen, \& van den Oord(1986)]{Mewe2} Mewe, R., Lemen,
J.R., and van den Oord, G.H.J. 1986, A\&AS, 65, 511

\bibitem[Montes et al.(2000)]{Montes00} Montes, M. J., Weiler, K. W.,
Van Dyk, S. D.; Panagia, N., Lacey, C. K., Sramek, R. A.; Park,
R. 2000, ApJ, 532, 1124

\bibitem[Mukai(1993)]{Mukai93} Mukai, K. 1993, Legacy, 3, 21
(available at http://heasarc.gsfc.nasa.gov/docs/journal/journals.html)

\bibitem[Nymark \& Fransson(2003)]{NF03} Nymark, T. K. \& Fransson,
C. (2003), in From Twilight to Highlight: The Physics of Supernovae,
eds. W. Hillebrandt \& B. Leibundgut (Berlin: Springer), 315

\bibitem[Petre et al.(1994)]{Petre94} Petre, R., Okada, K., Mihara,
T., Makishima, K., \& Colbert, E. J. M. 1994 PASJ, 46, L115

\bibitem[Plucinsky et al.(2003)]{Plu03} Plucinsky, P. P. et al. 2003,
X-Ray and Gamma-Ray Telescopes and Instruments for Astronomy,
ed. J. E. Tr\"{u}mper \& H. D. Tananbaum, SPIE, vol 4851, pp89 (13
co-authors)

\bibitem[Predehl \& Schmitt(1995)]{PS95} Predehl, P. \& Schmitt,
J. H. M. M. 1995, A\&A, 293, 889

\bibitem[Ryder et al.(2003)]{Ryd04} Ryder, S., et al. 2003, in preparation

\bibitem[Ryder et al.(1993)]{Ryder93} Ryder, S., Staveley-Smith, L.,
Dopita, M., Petre, R., Colbert, E. J. M., Malin, D., \& Schlegel,
E. M. 1993, ApJ, 416, 167 (R93)

\bibitem[Sambruna et al.(2000)]{SCU00} Sambruna, R. M., Chou, L. L.,
\& Urry, C. M. 2000, ApJ, 533, 650

\bibitem[Schlegel, Finkbeiner, \& Davis(1998)]{SFD98} Schlegel, D. J.,
Finkbeiner, D., \& Davis, M. 1998, ApJ, 500, 525
 
\bibitem[Schlegel et al.(1999)]{Schl99} Schlegel, E. M., Ryder, S.,
Staveley-Smith, L., Petre, R., Colbert, E. J. M., Dopita, M., \&
Campbell-Wilson, D., 1999, AJ, 118, 2689 (S99)

\bibitem[Schlegel, Petre, \& Colbert(1996)]{Schl96} Schlegel, Eric M.,
Petre, R., \& Colbert, E. J. M. 1996, ApJ, 456, 187

\bibitem[Schlegel(1995)]{S95} Schlegel, Eric M., Rep. Prog. Phys., 58, 1375

\bibitem[Swartz et al.(2003)]{Swartz03} Swartz, D. A., Ghosh, K. K.,
McCollough, M. L., Pannuti, T. G., Tennant, A. F., \& Wu, K. 2003,
ApJS, 144, 213

\bibitem[Weiler et al.(1996)]{Weiler96} Weiler, K. W., van Dyk, S. D.,
Sramek, R. A., \& Panagia, N. 1996, in Radio emission from the stars
and the sun, ed. A. R. Taylor \& J. M. Paredes (San Francisco:
Astronomical Society of the Pacific), 141

\bibitem[Zimmermann \& Aschenbach(2003)]{Zim03} Zimmermann, H.-U. \&
Aschenbach, B. 2003, A\&A, 406, 969 (astro-ph/0304322)

\end{thebibliography}
\end{document}